\documentclass{article}

\usepackage{blindtext}
\usepackage{hyperref}
\usepackage{graphicx} 
\usepackage{multirow} 
\usepackage{cite}
\usepackage{eucal}
\usepackage{todonotes}
\usepackage{amsmath,amssymb}
\usepackage{tikz}
\usepackage{caption}
\usepackage{subcaption}
\usepackage{xcolor}

\title{A photonic integrated circuit with reconfigurable nonreciprocal transmission and all-optical functionalities}

\author{Ang Li$^{1,2}$, Wim Bogaerts$^{1,2,3}$}

\begin{document}

\maketitle

\begin{enumerate}
 \item Photonics Research Group, Ghent university-IMEC, Department of Information Technology, Ghent University, Ghent 9052, Belgium
 \item Center for Nano- and Biophotonics, Ghent university, Ghent 9052, Belgium
 \item email: wim.bogaerts@ugent.be
\end{enumerate}

\begin{abstract}
We present a photonics integrated circuit on silicon substrate with reconfigurable nonreciprocal transmission that exhibits a large isolation ratio and low insertion loss. It also offers ability for all-optical functionalities, like optical computing gates, or a flip-flop. The circuit is a mixed cavity system of which the linear transmission can be tuned as a Fano resonance or electromagnetically induced transparency (EIT) using  two integrated heaters.  With high optical intensity inside the cavity, the Fano resonance and EIT peak induce a strong distortion due to thermal nonlinearities in the cavity, and these distortions depend on the transmission direction due to the asymmetric power distribution in the cavities. The resulting large isolation ratio is attributed to the inherent sharp slope of the Fano resonance and the large extinction ratio of the EIT peak. Thus, a high-power forward-propagating signal will trigger the nonreciprocal phenomenon for low-loss transmission, while backward transmission will see high loss irrespective of its power level, which is an outstanding improvement upon previously reported nonlinearity induced nonreciprocity in silicon photonics. The reconfigurability of the high transmission direction comes from the efficient control of the mode excitation and coupling inside the cavity using the integrated heaters.  Also, by using a separate pump laser, the device could be developed for all-optical functions like switching, logic and computing.
\end{abstract}

Nonreciprocal behavior in \textit{photonics integrated circuits} (PICs) is of fundamental interest and importance for on-chip isolators, circulators, signal processing, optical computing, and all-optical logic\cite{fan2012comment,jalas2013and}. However, the constraints of the Lorentz reciprocity theorem make it impossible to generate such a behavior in a linear, nonmagnetic and time-independent medium, which is the case for most PIC platforms. Accordingly, efforts to demonstrate nonreciprocal transmission in PICs rely upon breaking the time-invariance of the medium, using magneto-optic materials that shows direction-dependent permittivity tensors and introducing nonlinearity into the system. The first approach proves to be a promising engineering solution towards complete on-chip isolation, but it typically requires electro-optical modulation using complicated accompanying electronics and introduces significant power consumption. Moreover, it can induce unwanted frequency mixing due to the sidebands caused by the modulation \cite{yu2009complete, estep2014magnetic, sounas2017non, fan2018nonreciprocal}. 
Magneto-optic materials (e.g. as a waveguide cladding) in integrated optical structures (typically ring resonators) have also been demonstrated for nonreciprocal transmission\cite{tien2011silicon,ghosh2012yig, bi2011chip,shoji2014magneto,huang2017dynamically}. However, the path towards usable on-chip isolators is nontrivial due to the engineering difficulties of integrating those magneto-optic materials into established photonic integrated circuit (PIC) platforms. The third approach is to introduce asymmetric nonlinear effects like Brillouin induced transparency, optomechanically induced transparency, nonreciprocal Kerr effect in a silica, , thermal nonlinearities in asymmetric ring resonators or PT symmetric devices \cite{dong2015brillouin,kim2015non,shen2016experimental,del2018microresonator,sounas2017non,ruesink2018optical,fan2012all,mahmoud2015all,peng2014parity}. Our approach, which we report in this manuscript,  belongs in this last category. Our approach has the drawback that its nonreciprocal effects are dependent on the optical intensity, and thus might not be the perfect method for applications like optical isolators or circulators\cite{fan2012comment,shi2015limitations,fan2018nonreciprocal}. However, they have great value for other applications including nonlinear optical signal processing, logical gates or memory functions such as buffers or flip-flops.

\begin{figure}[htbp!]
    \centering
    \includegraphics[width=0.99\textwidth]{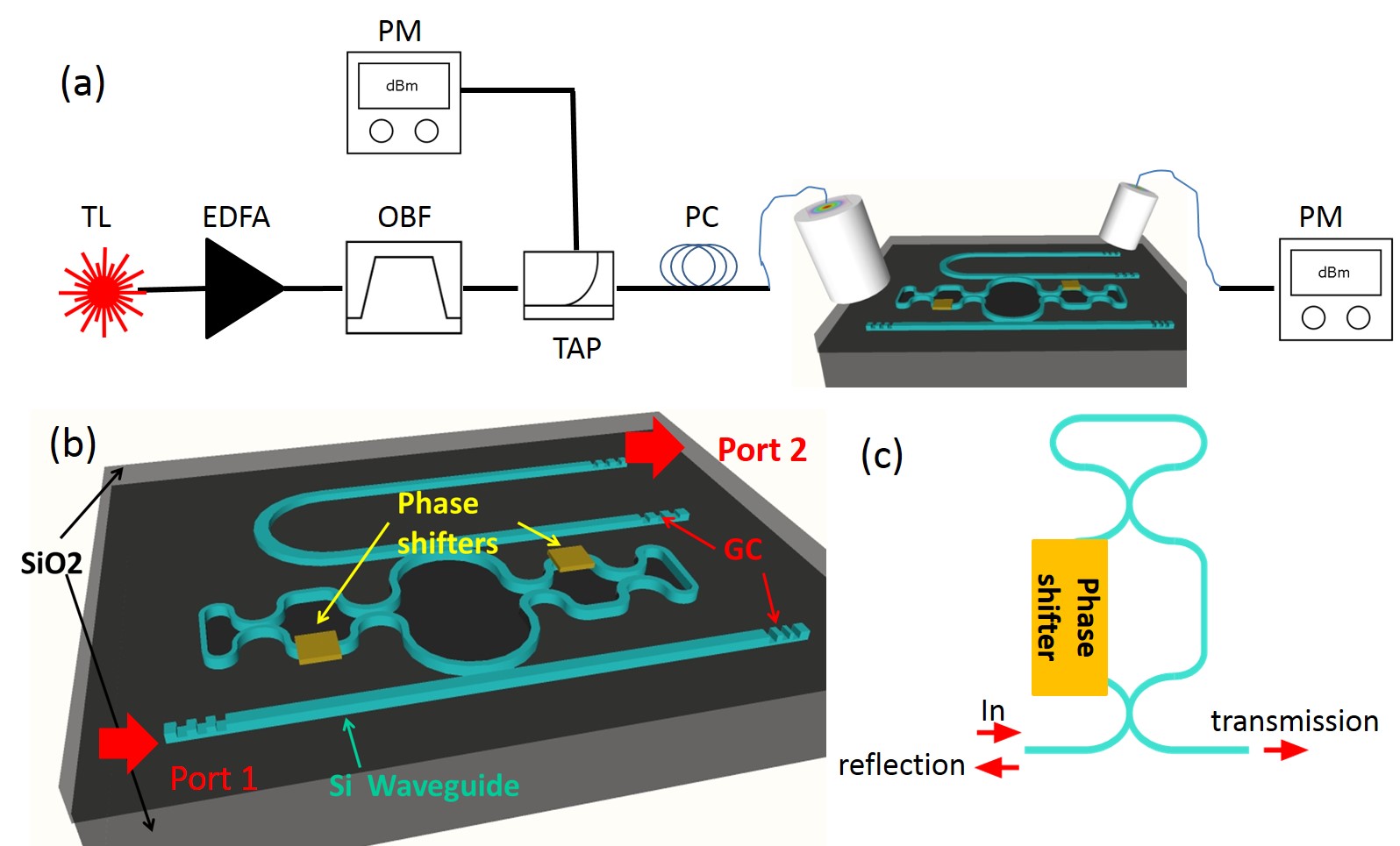}
    \caption{(a) The setup to chracterize the linear behavior of the device under test (DUT). TL: tunable laser. OBF: optical bandpass filter. Tap: a 1/99 splitter. PM: powermeter. PC: polarization controller. (b) The schematic of the DUT. GC: grating coupler. It is integrated on a silicon-on-insulator substrate. (c) the schematic of the tunable reflector in the DUT. }
    \label{fig:schematic}
\end{figure}

\par In this manuscript, we report a novel observation of nonreciprocal transmission in a mixed cavity system triggered by thermally induced nonlinearities at high optical input intensity and the structure could be implemented to other platform that shows intensity induced nonlinearity like Kerr effect. Similar with the methods based on optomechanically induced transparency and the Bruillouin induced transparency~\cite{dong2015brillouin,kim2015non, shen2016experimental,ruesink2018optical}, our method also involves a type of induced transparency, which is an optical analogue of electromagnetically induced transparency (EIT),  a phenomenon originating in atom physics\cite{fleischhauer2005electromagnetically}. Demonstrating EIT in (integrated) optics has already attracted significant interest as it is one of the most promising techniques to implement slow light structures for optical buffers or storage\cite{totsuka2007slow,lukin2001controlling,xu2006experimental,yang2009all,li2017tunable}. Our circuit further takes advantage of this to realize a reconfigurable nonreciprocal transmission with ultra high isolation ratio and low insertion loss. Briefly speaking, in the nonlinear regime (high input intensity), the transparency will only be present for one transmission direction and will be suppressed for light propagating in the opposite direction, thus resulting in a nonreciprocal transmission. While demonstrations of this principle using Brillouin scattering and/or optomechanics require customized optical chip implementations, our circuit is implemented as an electrically tunable photonic circuit on a standardized, publicly available silicon photonics technology platform and could be potentially transplanted to any other integration platform that supports similar nonlinearities. While compared with other approaches in silicon PICs, like the use of cascaded asymmetric ring resonators with thermal nonlinearities, our device shows a much smaller insertion loss as well as higher isolation ratio, and more importantly it can be reconfigured by tuning the integrated heaters, which means that we could 'program' the direction of high transmission. Also, our circuit only requires the forward signal to have high power in order to trigger the high transmission, while for the backward signal, no matter the power level, the transmission is always low. Besides triggering the nonreciprocal behavior with high input intensity, a separate pump laser could also be employed to release the requirement on the input intensity. Various all-optical functions could be built on this phenomenon. Even though this circuit might not be a perfect candidate for applications like an optical isolator (due to the power dependency performance) it could offer great value for applications like optical logic, computing and signal processing, where silicon photonics is emerging as a promising platform. \cite{fujii2006non, koos2009all,phang2015versatile, bernier2017nonreciprocal,li2017controlling, lecocq2017nonreciprocal}

\begin{figure}[htbp!]
    \centering
    \includegraphics[width=0.99\textwidth]{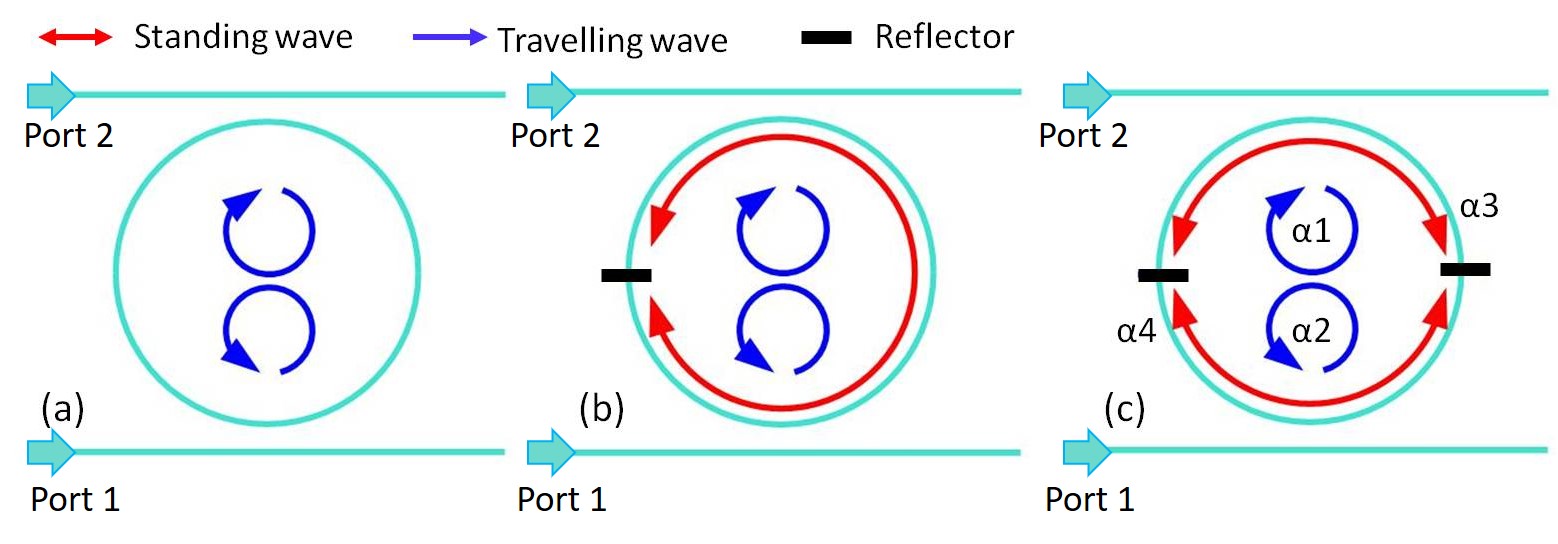}
    \caption{Resonant modes in (a) a pure ring resonator, (b) a ring resonator with a single internal reflector and (c) with two internal reflectors. No matter light is injected into port 1 or port 2, (a) and (b) always have identical power distribution among their respective modes. While for (c), depending on the injection direction, the modes $\alpha_3$ and $\alpha_4$ could have very different intensity distribution. Thus the structure generates different transmission spectrum with respect to transmission directions.}
    \label{fig:modes}
\end{figure}

\section{Silicon Photonics Integrated Circuit}
The schematic of our circuit is drawn in Fig.~\ref{fig:schematic}. The main circuit is fabricated on a 200~mm  silicon-on-insulator wafer using CMOS technology at IMEC with two metal heaters as phase shifters fabricated in local facility\cite{dumon2009towards}. The waveguide layer is 220~nm thick, with a 2~$\mu m$ thick buried oxide. the circuit consists of a ring resonator with two tunable reflectors inside. Each tunable reflector itself is a subcircuit consisting of an Mach-Zenhder-Inerferometer (MZI) and a waveguide that connects its two outputs, forming a loop mirror\cite{li2017experimental,li2018backcoupling}. By tuning the metal heater on top of one arm of the MZI reflector, its power reflectivity could be tuned efficiently, with only $0.5\pi$ needed to change the reflectivity from 0 to almost 100\%\cite{li2016design}. 

The characterization of the device for low optical input power is performed using setup shown in Fig.~\ref{fig:schematic}. An Agilent 8163B tunable laser with 1~pm wavelength step size is used as the source. The output will be first amplified by an erbium-doped fiber amplifier (EDFA) followed by a tunable optical filter to filter the noise of the EDFA. Then a 1/99 tap is used to split the input light to a HP power meter and the device under test (DUT). The power meter helps to monitor the exact input power to the DUT. Grating couplers (GC) are used to couple light from fiber to the chip and vice versa\cite{vermeulen2012reflectionless}. Each GC is expected to introduce around 6dB insertion loss near the measurement wavelength, based on separate measurements of reference circuits. The output of the DUT is connected to the Agilent photodetector to record the received power. Other equipment missing in this figure include two Keithley 2400 source meters used to control the two metal heaters as well as a temperature controller underneath the photonics chip to stabilize the ambient temperature of the chip.

\begin{figure}[htbp!]
    \centering
    \includegraphics[width=0.99\textwidth]{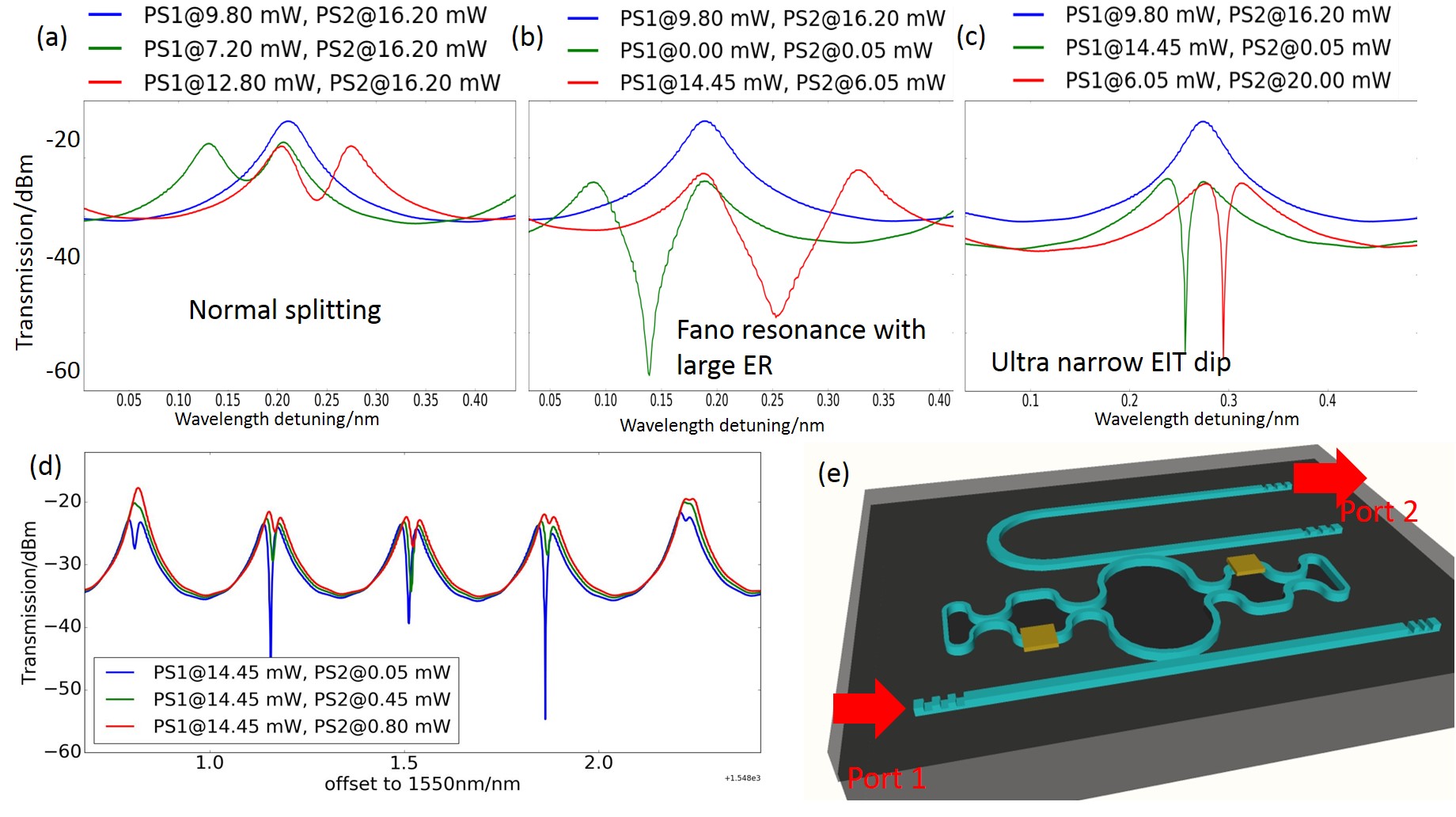}
    \caption{When the input power is low such that no nonlinearities inside the cavity are triggered, there would be four kinds of resonances at its output depending on the tuning conditions of the two phase shifters. (a) Lorentzian resonance with or without splitting. (b) Fano resonance with sharp slope. (c) ultra narrow and deep EIT dip. (d) shows how sensitive the EIT dip is to the phase shifter. Note that, in (a-c) the second peak could happen at either left or right side of the original peak. (e) re-plots the DUT and the transmission direction (from port 1 to port 2) under this measurement.
    }
    \label{fig:low_input}
\end{figure}

The linear behavior of the device has been reported in detail in \cite{li2017actively,li2017tunable}. But it's also worth briefly introducing here: when the corresponding power at the input of the DUT is as low as -5~dBm so that no nonlinearities are present, four different resonances at the drop port could be generated depending on the tuning conditions of the two metal heaters:
\begin{itemize}
    \item Lorentzian resonance (Fig.~\ref{fig:low_input}(a)) when both reflectors introduce zero reflections. In this case, there would be only one corresponding travelling wave mode inside the cavity at a given input port as shown in Fig.~\ref{fig:modes}(a). The peak transmission shows about 0.9~dB insertion loss, indicating the ring is critically coupled.
    \item Resonance splitting (Fig.~\ref{fig:low_input}(a)), when one reflector starts to show reflectivity. This reflection will couple the two circulating modes and lift their degeneracy\cite{li2016backscattering}. Thus they are resonant at slightly different frequencies but with identical intensity inside the resonator at steady state. Also a standing wave would exist due to the internal reflector as plotted in Fig.~\ref{fig:modes}(b). The splitting distance is proportional to the reflectivity, and it reaches the maximum at 0.5$\times$FSR when the reflectivity grows to 100\%\cite{li2017fundamental}. This could be interpreted in another way: under such circumstances, the two travelling waves would disappear with only the standing wave left. Since the standing wave cavity has double optical length compared to the ring resonator, so the FSR of the standing wave is half of that of the travelling wave cavity.
    \item Fano resonance (Fig.~\ref{fig:low_input}(b)), when both reflectors introduce reflections. Now the two reflectors would form two Fabry-Perot cavities with standing modes as illustrated in Fig.~\ref{fig:modes}(c). It has been demonstrated that the interaction of a discrete mode (high $Q$) with a smooth background mode (low $Q$) would generate a Fano resonance when there is frequency detuning~\cite{fano1961effects,fan2002sharp,luk2010fano,li2017actively}. In our circuit, the two standing modes of the low $Q$ FP cavities serve as the background modes for the discrete modes corresponding to the ring resonances. But the Fano resonance pattern has a qualitative difference compared to previously reported Fano resonances~\cite{qiu2012asymmetric,li2012experimental,zhang2016optically,zhao2016tunable}, as in our circuit, the ring resonance is not a single Lorentzian resonance, but instead shows splitting, which can be modeled by two closely spaced Lorentzian resonances. Thus it actually generates a double-Fano pattern.
    \item EIT (Fig.~\ref{fig:low_input}(c)). Tightly linked to the appearance of the Fano resonance, when the low $Q$ FP mode and the high $Q$ resonance splitting have zero frequency detuning, a phenomenon called EIT will be triggered~\cite{fleischhauer2005electromagnetically, li2017tunable}, with ultra narrow bandwidth and usually large extinction ratio. This is the key for our device to exhibit large isolation ratio and low insertion loss at the nonreciprocal transmission. The frequency detuning between the FP modes and the ring resonances could be tuned by controlling the two reflectors. Note in Fig.~\ref{fig:low_input}(d), only a tiny amount of electrical tuning is needed in order to change the resonance pattern from Fano to EIT, which confirms that it's the frequency detuning that matters as this tiny amount of power is negligible to change the reflectivity ($Q$-factor of the FP modes).
\end{itemize}
One important feature to highlight here: for the resonance splitting, Fano resonance and EIT, the second peak can appear either blueshifted or redshifter with respect to the original peak depending on the operating conditions of the two phase shifters as evident in Fig.~\ref{fig:low_input}(a-c), which is the key feature to realize re-configurable nonreciprocal transmission. 

\begin{figure}[htbp!]
    \centering
    \includegraphics[width=0.99\textwidth]{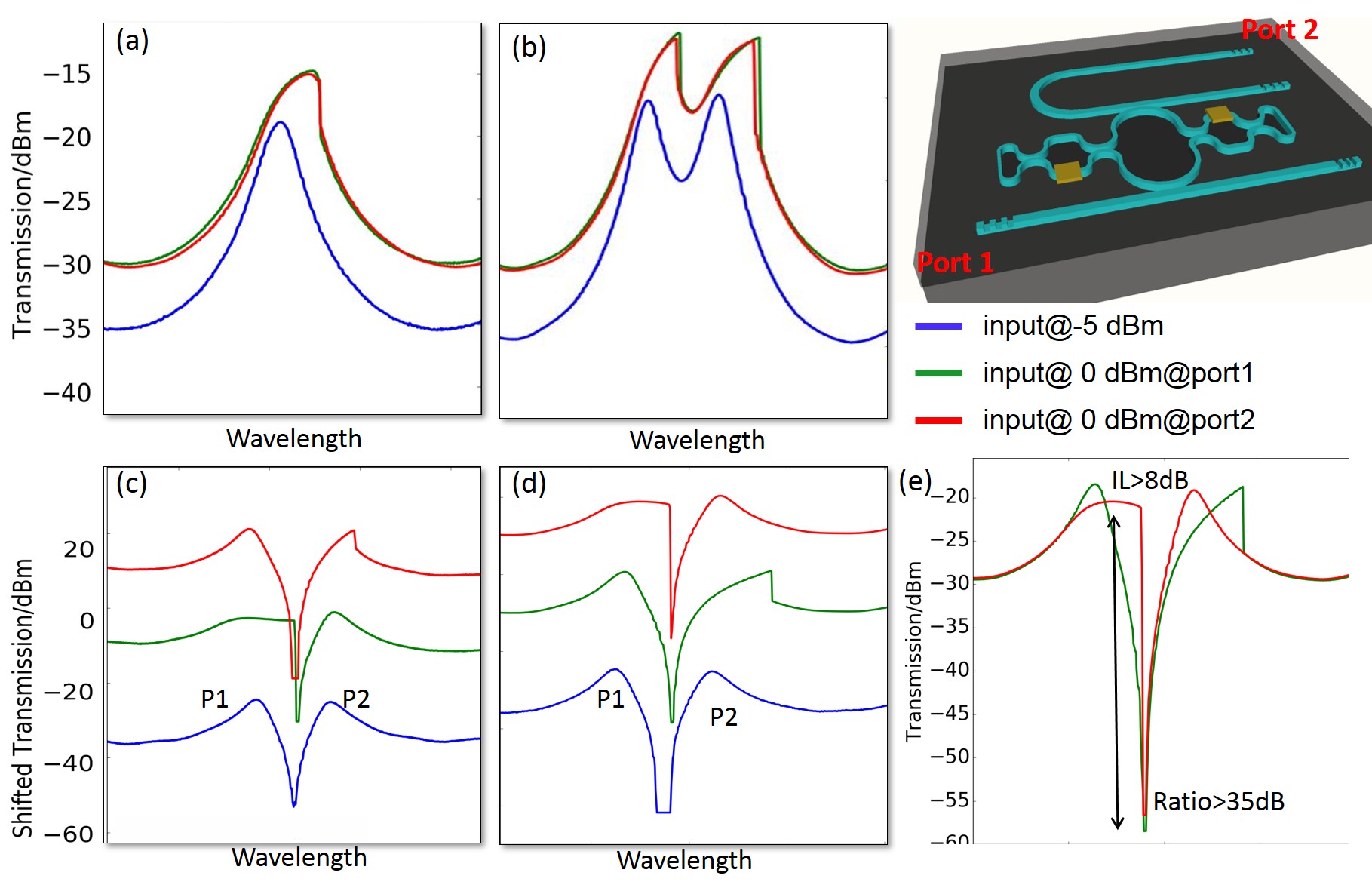}
    \caption{When the input power is high, nonlinearity-induced distortion will emerge for all types of resonances. Green curve represents the transmission from port 1 to port 2 at high input power, while red curves shows the reverse transmission at high input power. Blue curve refers to the transmission from port 1 to port 2 at low input power (without nonlinearities). (a) and (b) confirm that the spectra are identical at the Lorentzian resonance and the resonance splitting case, irrespective of the transmission direction. While distortions of the Fano resonances (c-d) become dependent on transmission-direction and it could be reconfigured, thus nonreciprocal transmission is generated (e). In detail, in (c), transmission from port 1 to port 2 shows distortion at the left peak, while it's the right peak that shows distortion at the reverse transmission direction (from port 2 to port 1).  This leads to a high transmission from port 2 to port 1 and low transmission reversely at specific wavelength (as shown in (e)). But in (d), the peak that shows distortion at each transmission direction is swapped, leading to high transmission from port 1 to port 2 at the same wavelength.
    }
    \label{fig:nl}
\end{figure}

\section{high power input}
When the input power is increased to a certain value (over 0~dBm at the input of the DUT), thermal nonlinearities in the cavity will be induced around the resonance wavelength. The resonance spectrum will show distortion and will red-shift compared to the undistorted resonance peak in the linear regime, as evident in Fig.~\ref{fig:nl},  because the strong field inside the resonator will induce self-heating which in turn changes the refractive index of the silicon waveguide~\cite{almeida2004optical,priem2005optical,xu2006carrier, fan2012all}. At the Lorentzian resonance or resonance splitting regime, this distortion is independent of the transmission directions (Fig.~\ref{fig:nl}(a-b)), which means the spectrum is identical irrespective of the transmission direction, as all the modes inside the cavity have identical intensity distributions irrespective of the transmission direction as illustrated in Fig.~\ref{fig:modes}(a-b), and therefore the entire system is symmetric and reciprocal. 
\par Interestingly, the distortion of the Fano resonance becomes dependent on the transmission direction. The Fano resonance at the nonlinear regime is plotted in Fig.~\ref{fig:nl}(c-d) (the resonances are shifted vertically with spacing of 20~dB in order to provide a clear comparison). We define the transmission from port 1 to port 2 as $T_{12}$  and vice versa. The left peak of the resonance is defined as $P_1$ and the right one as $P_2$. The feature is that either $P_1$ or $P_2$ shows that the distortion is transmission dependent. This is due to the asymmetric power distribution among the two standing wave modes ($\alpha_3, \alpha_4$, see Fig.~\ref{fig:modes}) determined by the reflectivity of the reflectors. If the reflector is working with strong reflectivity (for instance 90\%), then for the transmission from port 1 to port 2, most of the power will be confined at the lower FP cavity, while for backward transmission (from port 2 to port 1) most of the power will be confined in the top FP cavity. This asymmetric field distribution leads to different peak that shows nonlinearity-induced distortion. In such a way, the behavior of the device is dependent on the transmission direction of the light, resulting in a nonreciprocal transmission with large isolation ratio as evident in Fig.~\ref{fig:nl}(e). And as we are able to determine which peak to distort by tuning the reflectors as evident in Fig.~\ref{fig:nl}(c-d), we can reconfigure the nonreciprocal transmission, in other words, we could determine which direction to show high transmission, and its corresponding reverse direction becomes low transmission. However, we notice that at the nonlinear Fano resonance regime, the insertion loss is relatively high (9dB) as splitting is present for both transmission directions, which reduces the overall peak transmission (Fig.~\ref{fig:nl}(e)).

\begin{figure}[htbp!]
    \centering
    \includegraphics[width=0.99\textwidth]{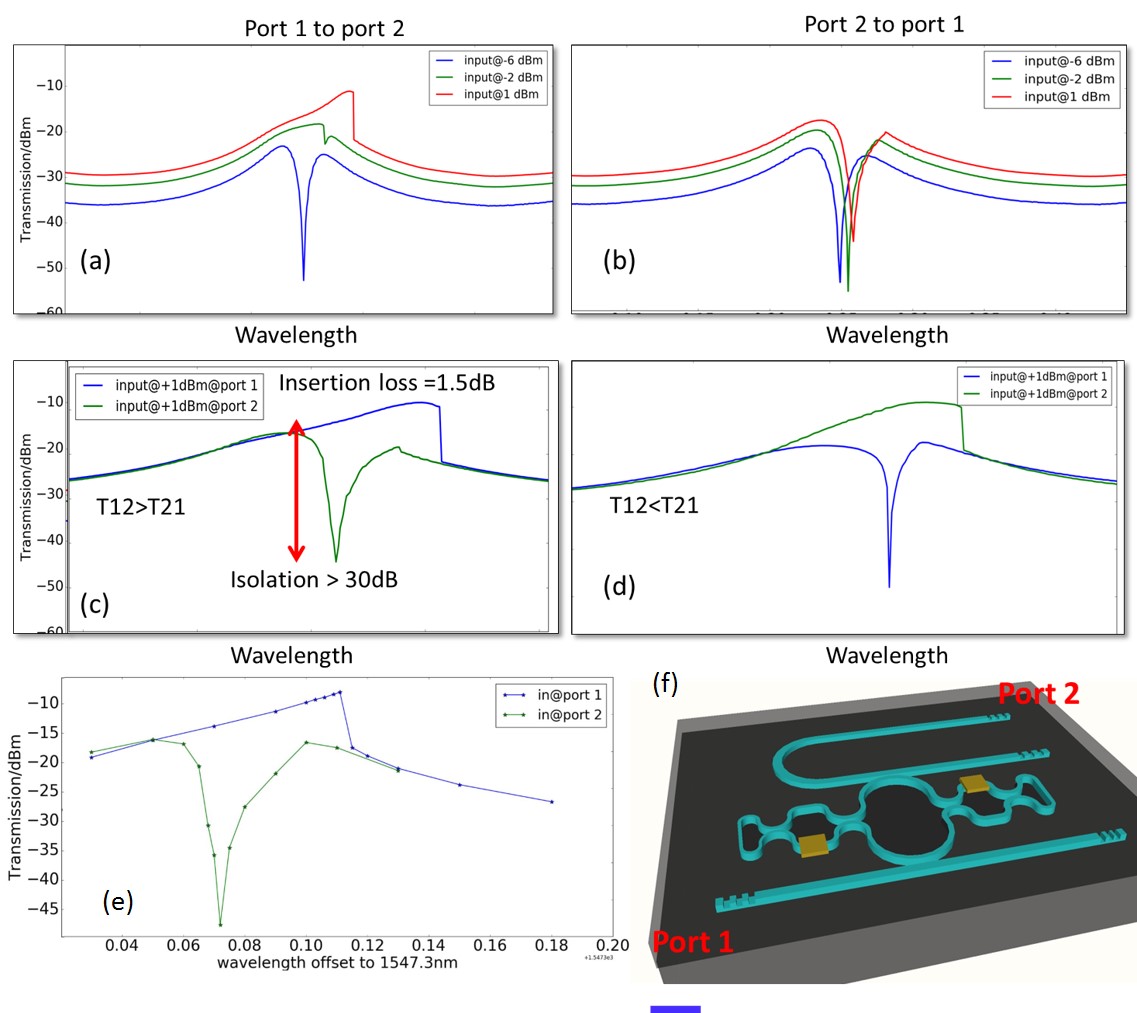}
\caption{At the EIT regime, resonance also shows distortion and it's dependent on the transmission direction. (a) shows the transmission spectra from port 1 to port 2 at varying input power. Note how the EIT peak evolves to a Lorentzian resonance and ends up with high transmission. (b) shows the spectra from port 2 to port 1 with varying input power and the EIT peak with low transmission is always present. This leads to nonreciprocal behavior with transmission from port 1 to port 2 as the high direction as plotted in (c). (d) plots the fact that the device could be reconfigured to another condition where transmission from port 2 to port 1 is higher than the opposite direction. (e) shows the manual sweep to confirm the nonreciprocal transmission. The insertion loss for the high transmission direction is as low as 1.5~dB and the isolation ratio is over 30~dB. Also, we could configure which transmission direction as the high transmission. 
}
    \label{fig:eit_nl}
\end{figure}

In order to keep the nonreciprocal transmission with both large isolation ratio and low insertion loss, the EIT phenomenon should be taken advantage of as plotted in Fig.~\ref{fig:eit_nl}. Nonlinearities still distort the EIT peak dependent on the transmission direction. For high transmission direction, the nonlinearities could eliminate the EIT peak, leading to a distorted Lorentzian resonance with high transmission as it is the left peak $P_1$ that shows nonlinearity, such that the two peaks come closer together until the resonance degenerates into a non-split Lorentzian resonance. Thus the insertion loss is dramatically reduced to only 1.5~dB as in Fig.~\ref{fig:eit_nl}. The insertion loss (IL) is directly dependent on the configuration of the ring resonator, including its propagation loss and coupling coefficients. While for the opposite transmission direction, it's the right peak $P_2$ that shows nonlinearity and red-shifting, so the EIT peak is still present with much lower transmission, resulting a high isolation ratio over 30~dB. This brings another advantage of our method compared to previous methods: for the backscatter wave, we do not require high power as the desired (low) transmission is the same in both the linear and nonlinear regime, because irrespective of the backscattered intensity, there is always an EIT dip with very low transmission. In addition, we are able to dynamically configure the device to have high transmission in either one of the two possible directions, and low transmission in the other one. 

\par The measurements discussed above were done by sweeping the input wavelength continuously from short to long wavelengths. Because for high input power the ring resonator will exhibit a bistable regime near the resonance\cite{priem2005optical,xu2006carrier}, the wavelength sweep will only show one branch of this bistable regime and then abruptly transition to the other branch.  To show that the nonreciprocal transmission is not dependent on this particular sweep direction, we perform a manual scan with fixed input power  at +1~dBm, where we turn off the laser for each step, then set the input wavelength and turn the laser back on to record the transmitted power, and then repeat this for each wavelength. The results are plotted in Fig.~\ref{fig:eit_nl}(e). The nonreciprocal pattern clearly persists.

\par This device with nonreciprocal transmission may not be suitable to function as a true optical isolator, but it would be of great interest to be used for nonlinear signal processing. For instance, it could be used as a optical comparator or signal re-generator that detects digitally modulated optical signal with reference to a threshold power. Power larger than this threshold is detected as “1” and lower as "0". The threshold could be set close to the trigger point of this nonlinear behavior (EIT$\rightarrow$ Lorentzian), then the "1" signal will end up with very high transmission with negligible loss (~1 dB) while the "0" signal will be considerably lost by the device. As a consequence, the extinction ratio of the digitally modulated signal will be dramatically boosted. Our device utilizes thermal nonlinearities, which act on a time scale of $\mu s$ and thus are not suitable for high speed data rates. However, the fundamental principle is self-phase modulation (optical intensity dependent refractive index). So the concept could be implemented in platforms with fast nonlinear effects like Kerr effect or free carrier dispersion in silicon\cite{wang2013theoretical}.

\begin{figure}[htbp!]
    \centering
    \includegraphics[width=0.99\textwidth]{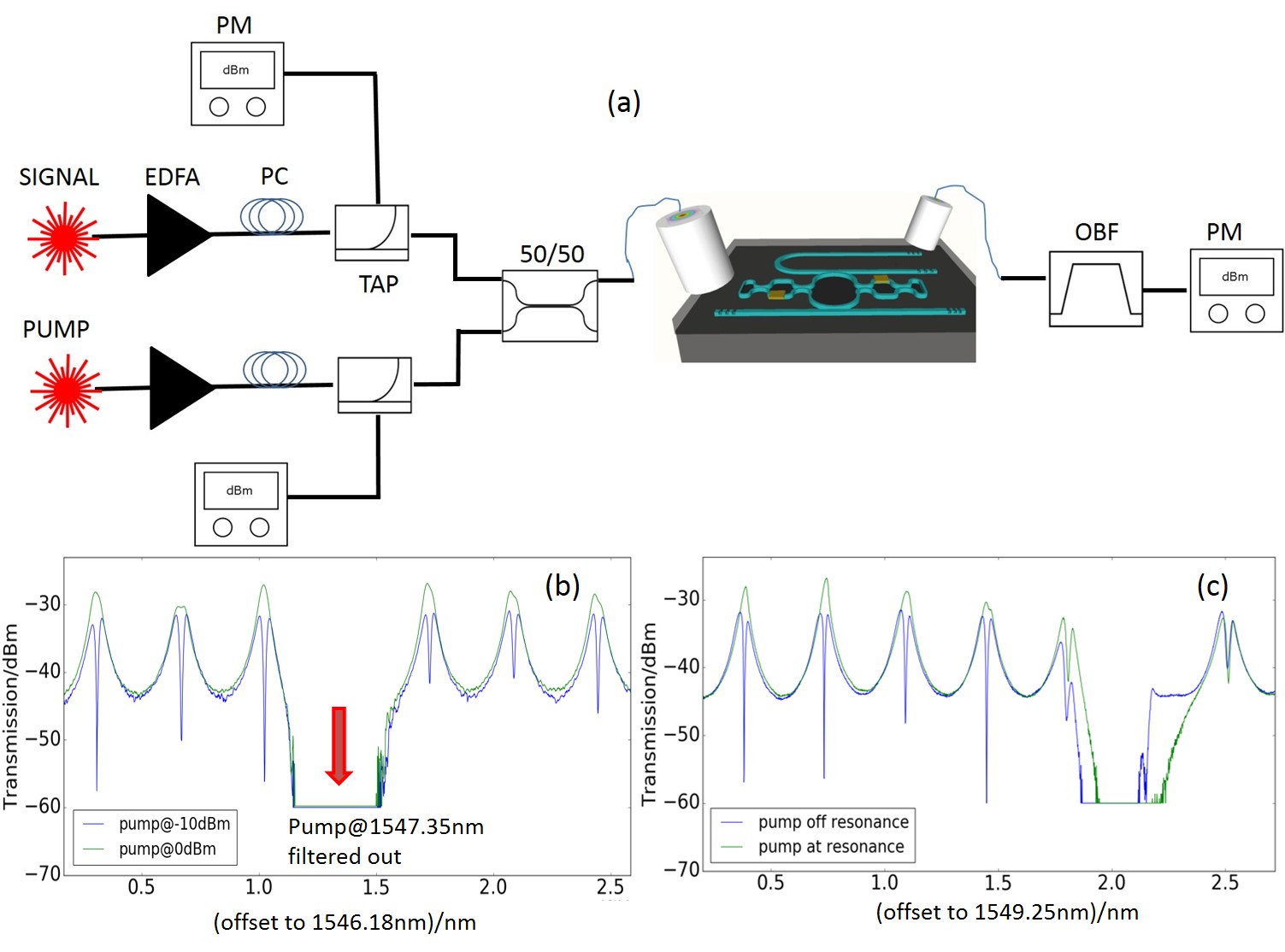}
    \caption{Pump-Signal measurement. (a) shows the measurement setup. A pump laser and a signal laser are mixed through a 50/50 coupler and fed into the port 1 of the DUT together. The pump laser is fixed at certain wavelength while the signal laser is swept at 1~pm step. (b) shows the spectra of the signal laser at different pump power with the pump wavelength aligned to one of the resonance waveguide. Clearly, when the pump power is high, the device generates high transmission Lorentzian resonance. It behaves like a switch to the device that control its output transmission. (c) plots the measured spectra at fixed pump power but with different pump wavelength. Only when the pump is at the resonance wavelength, the transition from EIT peak to Lorentzian resonance could happen.
    }
    \label{fig:setup2}
\end{figure}

\section{pump laser}

Up till now, we have described a mode of operation that required the input power of the forward direction to be high enough in order to the induce thermal nonlinearity in the ring resonator to generate high transmission. However, the nonreciprocal transmission can also be triggered using a separate pump laser to induce the asymmetric intensity distribution among the two standing wave modes. To characterize it, we use a Santec TL510 tunable laser source (C-band) as the separate pump laser and perform some measurements in a pump-probe setup. For all the measurements, the device is configured to allow high transmission from port 1 to port 2 and low transmission reversely. In the following section, with 'power' we mean the power at the input of the DUT.
\par First of all, the pump laser is tuned to a resonance wavelength (1547.35~nm) and it is coupled with the signal laser through a 50/50 coupler before being fed into port 1. The setup is in Fig.~\ref{fig:setup2}(a). By setting the signal power to be very low (-10~dBm), we plot the spectra of signal laser measured at port 2 with two pump levels as shown in Fig.~\ref{fig:setup2}(b). Clearly, at low pump level (-10~dBm), the spectrum shows an EIT pattern while when the pump levels increases to 0~dBm, the EIT pattern has disappeared and a single Lorentzian resonance is present as the pump is building up inside the cavity while making the desired changes to the cavity transmission. To further confirm the impact of the pump laser, we set the pump power at 0~dBm and measure the spectra at different pump wavelengths, one is at resonance (1551.09~nm) and the other one is off-resonance (1551.01~nm) as plotted in Fig.~\ref{fig:setup2}(c). When the pump laser is off resonance, the EIT pattern is still observed as now the pump laser could not accumulate inside the cavity. 
\par These two measurements confirm that using a separate pump laser, even for a weak input signal, the EIT pattern could be converted to a Lorentzian resonance and lead to high transmission. Then the device could support applications like all-optical switch, all-optical logic and nonlinear signal processing. For instance, the device could be used as an all-optical "AND" logic gate with two inputs (pump and signal). Either of them is high (1), the output of the device would be high (1). While the output would be low (0) in case both of them has low power (0). Another example is that, the device could be considered as a all-optical D type flip-flop, with the pump laser serving as the control signal, the transmission from input to output would only be allowed at high level of the control signal. Once again, even if our device is using thermal effects, the concept should also work with other fast nonlinear effects that make it more suitable for higher data rates.

\section{Conclusion}
In this manuscript, we propose a programmable photonics integrated circuit that is fully integrated onto silicon substrate with a compact footprint. It is a mixed cavity system supporting 4 modes inside the cavity. Using two metal heaters, the individual modes and the coupling between them could be efficiently controlled, so is the corresponding output of the system. Various resonance patterns have been observed and explained, including Lorentzian resonance, split resonance, Fano resonance and optical analogue of EIT. The behavior of the device at high input power that is able to induce thermal nonlinearity is characterized in detail and it's observed that at Lorentzian resonance or split resonance pattern, the device is reciprocal while at Fano resonance or EIT regimes, the device turns to nonreciprocal, due to the asymmetric intensity distribution among the modes inside the cavity. Specifically, at one transmission direction, the nonlinearity induced resonance distortion can eliminate the EIT transmission dip and result in a Lorentzian resonance with high transmission, while for the opposite direction, the EIT transmission dip remains, and therefore the circuit has a low transmission. This nonreciprocal transmission is accompanied with an ultra high isolation ratio and low insertion loss, which exhibits significant improvement compared to previously demonstrated nonreciprocal transmission in silicon PIC. More importantly, the high transmission direction could be reconfigured by tuning the metal heaters, making it a unique nonreciprocal PIC. Moreover, we experimentally demonstrate another all-optical approach to trigger the transition from an EIT pattern to a Lorentzian resonance, which is using a separate pump laser at one of the resonance wavelength to generate the necessary modes conditions inside the cavity. This circuit could form the basis of a variety of optical functions, including but not limited to all-optical switching, all-optical flip-flop, all-optical signal regeneration, and all-optical logic gates.






\end{document}